\documentclass[namedreferences]{SolarPhysics}
\usepackage[optionalrh,solaenum]{spr-sola-addons} % For Solar Physics
\usepackage{graphicx}        % For eps figures, newer & more powerfull
\usepackage{color}           % For color text: \color command
\usepackage{url}             % For breaking URLs easily trough lines
\newcommand{\beq}{\begin{equation}}
\newcommand{\eeq}{\end{equation}}
\newcommand{\bfig}{\begin{figure}}
\newcommand{\efig}{\end{figure}}
\newcommand{\ben}{\begin{enumerate}}
\newcommand{\een}{\end{enumerate}}

\begin{document}
\begin{article}
\begin{opening}

\title{The Force-Free Electrodynamics Method for the Extrapolation
of Coronal Magnetic Fields from Vector Magnetograms}

\author{I.~\surname{Contopoulos}}
\runningauthor{Contopoulos} \runningtitle{FFE in the Solar Corona}

   \institute{Research Center for Astronomy and Applied
     Mathematics (RCAAM), Academy of Athens,
     4 Soranou Efessiou Str., Athens 11527, Greece,
     email: \url{icontop@academyofathens.gr}}
\begin{abstract}
We present a new improved version of our force-free
electrodynamics (FFE) numerical code in spherical coordinates that
extrapolates the magnetic field in the inner solar corona from a
photospheric vector magnetogram. The code satisfies the
photospheric boundary condition and the condition $\nabla\cdot
{\bf B}=0$ to machine accuracy. The performance of our method is
evaluated with standard convergence parameters, and is found to be
comparable to that of other nonlinear force-free extrapolations.
%, and the force-free condition $\left<(\nabla\times {\bf
%B})\times {\bf B}\right>=0$ to about $10\%$ of $\left<B^2\right>/d$,
%where $d$ is the spatial resolution of the numerical grid.
\end{abstract}
\end{opening}
%-------------------------------------------------

%\newpage
\section{Introduction}
     \label{S-Introduction}
Force-free electrodynamics (hereafter FFE) is a formal name for
time-dependent electromagnetism in an ideal plasma with negligible
inertia and gas pressure ($\beta\approx 0$). The formalism of FFE
has been developed for various relativistic astrophysical
applications where the plasma supports electric currents and
electric charges (pulsars, astrophysical jets, gamma-ray bursts,
{\it etc.}). The equations of FFE are Maxwell's equations with
nonzero electric currents,
\begin{equation}
\frac{\partial {\bf E}}{\partial t} = c \nabla\times {\bf B} -
4\pi {\bf J} \ , \ \ \frac{\partial {\bf B}}{\partial t} = -c
\nabla\times {\bf E}\ , \label{Maxwell}
\end{equation}
complemented by the divergence-free, ideal MHD, and force-free
conditions
\begin{equation}
\nabla\cdot {\bf B}=0\ ,\ \ {\bf E}\cdot {\bf B} = 0\ ,\ \
\rho_{\rm e} {\bf E}+\frac{1}{c}{\bf J}\times {\bf B} = 0\ .
\label{forcefree}
\end{equation}
Here, ${\bf J}$ and $\rho_{\rm e}\equiv (4\pi)^{-1}\nabla\cdot
{\bf E}$ are the electric current and charge densities,
respectively. One can solve the above set of equations for ${\bf
J}$ and thus express the electric current density as a function of
the electric and magnetic fields, namely
\begin{equation}
{\bf J} = \frac{c}{4\pi}\nabla \cdot {\bf E}\ \frac{{\bf E}\times
{\bf B}}{B^2} + \frac{c}{4\pi}\frac{({\bf B}\cdot \nabla\times
{\bf B} - {\bf E}\cdot \nabla\times {\bf E})}{B^2}\ {\bf B}
\label{J}
\end{equation}
\cite{gruzinov99}. One can then numerically integrate Maxwell's
equations to follow the temporal evolution of any force-free
ideal-MHD system from an initial to a final configuration. In the
particular case when ${\bf
E}\stackrel{t\rightarrow\infty}{\longrightarrow} 0$, the final
configuration is (in general) a non-linear force-free
magnetostatic field, {\it i.e.} it satisfies the relations
\begin{equation}
{\bf J}= \frac{c}{4\pi}\nabla\times {\bf B}\ ,\ \ \nabla\cdot {\bf
B}=0\ ,\ \ {\bf J}\times {\bf B}=0\ .
\end{equation}

The FFE method has been applied successfully in two main
astrophysical applications; pulsars (\opencite{spitkovsky06};
\opencite{kc09}), and the solar corona (\opencite{CKG}, hereafter
CKG). The numerical codes developed for these problems are
Cartesian finite-difference time-domain (FDTD) codes. They are
based on the staggered mesh algorithm of \inlinecite{yee66} where
in every computational cell, electric field components are defined
in the middle of and along cell edges, whereas magnetic field
components are defined in the middle of and perpendicular to cell
faces. It is very important to notice that this prescription
satisfies the divergence-free condition $\nabla\cdot {\bf B}=0$ to
machine accuracy. \inlinecite{kc09} introduced also an outer
absorbing non-reflecting boundary condition in the form of a
perfectly matched layer (PML). CKG implemented the FFE method in
the study of the inner solar corona by evolving the photospheric
boundary condition toward a given distribution for the radial
component of the magnetic field. This is achieved with the
introduction of photospheric electric fields which gradually die
out as the required photospheric condition is approached. During
that evolution, the FFE equations are numerically solved in the
solar corona, and in the limit when electric fields die out
everywhere, a non-linear force-free magnetostatic configuration
emerges.

There are several problems with this approach: a) The method is
based on knowledge only of the radial component of the
photospheric magnetic field; therefore the final solution is not
unique. Different initial conditions and different photospheric
evolutionary paths yield different magnetostatic configurations.
b) Numerical dissipation cannot be avoided, and therefore, the
longer a particular implementation takes to converge to a certain
acceptable accuracy, the smaller the final remaining electric
currents in the corona, and the closer the final solution
approaches the potential (current-free) equilibrium. c) The
numerical grid is Cartesian. This works well around the poles, but
is unsuitable to appropriately model the solar photosphere in
general.

For all of the above reasons, we decided to improve our code, and
to incorporate the photospheric boundary condition supplied from a
full vector magnetogram .

\section{Implementation of the Vector Magnetogram Boundary Condition}

The new improved version of our code is written in the natural
coordinates for the solar corona, namely heliocentric spherical
ones $(r,\theta,\phi)$. In order to satisfy the photospheric
boundary condition exactly, we altogether skip the photospheric
relaxation algorithm described in CKG, and begin instead with the
following `sea urchin'-like initial magnetic field configuration:
\[
B_r(r,\theta,\phi; t=0)={\cal B}_r(\theta,\phi)(r_{\odot}/r)^2\ ,
\]
\[
B_\theta(r_{\odot},\theta,\phi;t=0)={\cal B}_\theta(\theta,\phi) \
, \]
\[
B_\phi(r_{\odot},\theta,\phi;t=0)={\cal B}_\phi(\theta,\phi)\ ,\ \
\mbox{and}
\]
\[
B_\theta(r>r_{\odot},\theta,\phi;t=0)=B_\phi(r>
r_{\odot},\theta,\phi;t=0)=0\ .
\]
This configuration is obviously divergence-free, and satisfies the
photospheric boundary condition ${\bf
B}(r_\odot,\theta,\phi)\equiv {\bf {\cal B}}(\theta,\phi)$
exactly. We also set ${\bf E}=0$ on the photosphere at all times.
Notice that the algorithm that we implement (\opencite{yee66})
satisfies $\nabla\cdot {\bf B}=0$ to machine accuracy {\em only if
the initial configuration is divergence-free}. In other words, the
algorithm does not implement the divergence-free condition. It
inherits and preserves it. Notice also that the initial
configuration is filled with electric currents in the $\theta$-
and $\phi$--directions, but not in the radial one.

We then proceed as in CKG with a numerical integration of
Equations~(\ref{Maxwell}) and (\ref{J}). Initially, the magnetic
field configuration becomes torn through numerical reconnection.
The sea urchin-like configuration gradually disappears. Transient
electromagnetic-type waves travel through the integration volume
and when they reach the outer radial boundary they are absorbed by
the PML (see below). Eventually, this transient activity dies out,
electric fields are observed to converge to ${\bf
E}\stackrel{t\rightarrow\infty}{\longrightarrow} 0$, and a
non-linear force-free magnetostatic equilibrium is gradually
reached. Notice that during this evolution, the divergence-free
and vector photospheric boundary conditions are everywhere
satisfied.

Our method can be implemented in both global and local magnetic
field extrapolations. In global extrapolations, the numerical
integration volume lies between an inner radius $r_{\rm
in}=r_{\odot}$ (the photosphere), and an outer radius $r_{\rm
out}$ that corresponds to the distance beyond which the solar wind
becomes dynamically important. The integration volume corresponds
to the lower solar corona where we assume that force-free
magnetostatic conditions apply to a certain extent. In the test
simulations presented in the next section, we took $r_{\rm
out}=2r_\odot$. In general, $r_{\rm out}$ may vary between $2
r_{\odot}$ and $3 r_{\odot}$, depending on the strength of the
solar wind. Beyond that radius, we implement a PML non-reflecting
absorbing layer of thickness $0.5 r_{\odot}$. In the polar
direction, the simulation is limited between a minimum and a
maximum polar angle $\theta_{\rm min}= 20^\circ$ and $\theta_{\rm
max}= 160^\circ$, in order to avoid the polar singularity of the
spherical coordinate system. In realistic solar coronal magnetic
field configurations, the polar regions are covered by coronal
holes, and therefore, their exclusion does not significantly
affect the extrapolation. Finally, in the azimuthal direction we
implement periodic boundary conditions. The test simulations
presented in the next section are all global extrapolations except
for the last one that is limited in the azimuthal extent.

\section{Evaluation of Numerical Code}

In order to evaluate the performance of our code, we implement
vector photospheric boundary conditions derived from standard
solutions that are used as benchmarks, and evaluate the
convergence of the extrapolation based on standard convergence
parameters introduced in \inlinecite{Wetal}, \inlinecite{Setal},
and \inlinecite{Aetal} (see the Appendix for their detailed
expressions). The benchmark solutions are known either
analytically (dipole), or quasi-analytically (\opencite{LL90},
hereafter LL). In most previous implementations, benchmark
boundary conditions on the $xy$-plane were obtained by displacing
the LL solution some distance $l$ below the $xy$-plane, and by
rotating it by an angle $\Phi$ with respect to the $y$-axis (see
the original LL paper for details). To the best of our knowledge,
the only previous work where the LL solutions were applied in a
coronal magnetic field configuration over a curved photospheric
boundary is \inlinecite{Tadesse}. We thus decided, for the sake of
comparison, to also implement their boundary condition. We
investigated the following test cases:
\begin{itemize}
\item Case DF: a dipole magnetic field off-axis by $0.5 r_{\odot}$
\item Case LL11: LL solution with $n=1$, $m=1$, $l=0.25r_{\odot}$
(displacement below the $xy$-plane), $\Phi=\pi/10$  (as in
\opencite{Tadesse}) \item Case LL13: LL solution with $n=1$,
$m=3$, $l=0.1 r_{\odot}$ (displacement below the $yz$-plane),
$\Phi=4\pi/5$ \item Case LL31: LL solution with $n=3$, $m=1$,
$l=0.1 r_{\odot}$ (displacement below the $yz$-plane),
$\Phi=4\pi/5$
\end{itemize}

Case~DF is current-free (potential), but the LL solutions are not.
The radial resolution of these particular simulations is $0.1
r_\odot$, and the angular resolution $4^\circ\times 4^\circ$
(heliocentric). We set $\theta_{\rm min}=20^\circ$ and
$\theta_{\rm max}=160^\circ$. Our computational grid in
($r,\theta,\phi$) has a resolution of $10\times 34 \times 90$.

In Table~1, we plot the values of our convergence parameters in a
subregion of our computational volume (we chose
$r_\odot<r<1.5r_\odot$, $30^\circ < \theta<60^\circ$, $30^\circ <
\phi< 60^\circ$) in the force-free extrapolations obtained for the
above four benchmark solutions. The first four parameters
characterize the convergence to the benchmark solution (unity
corresponds to a perfect match). The last two parameters
characterize how well the extrapolation satisfies the force-free
and divergence-free conditions respectively. In all cases, the
divergence-free and boundary conditions are satisfied to machine
accuracy. The force-free condition is also very important for us.
We evolve the numerical simulation till we obtain a minimum of the
parameter $\mbox{CWsin}$ at a level roughly below 0.15. This
guarantees that the solutions that we present are indeed
non-linear force-free magnetostatic configurations that satisfy
the given vector magnetogram boundary conditions. Notice that the
force-free parameter (Equation~(\ref{ff})) does not apply in
potential extrapolations where $J\rightarrow 0$.

The performance of our code is satisfactory, which suggests that
the present numerical method is promising. In some benchmark
cases, it outperforms existing Cartesian grid extrapolations (see
{\it e.g.} \opencite{JF12}), and is certainly on a par with the
spherical grid extrapolation of \inlinecite{Tadesse} (compare our
case LL11 with Case 1 in their low resolution grid). In some
cases, though, the resulting extrapolation does not converge to
the benchmark solution in the volume under investigation, as
manifested for example in case LL13 where two convergence
parameters differ significantly from unity. We offer two possible
explanations for this result: a) An extrapolation where a
significant amount of magnetic flux crosses the outer boundaries
of the numerical integration region is influenced by the presence
of those boundaries more than another extrapolation where most
magnetic field lines are contained within the computational
volume. b) A non-linear extrapolation is not unique when current
sheets are allowed to develop in the solution. This needs to be
confirmed with more detailed evolutionary full MHD extrapolations.
Notice that our force-free code handles current sheets as
tangential magnetic field discontinuities with $|B|$ continuous
everywhere (see \inlinecite{CKF} and \inlinecite {kc09} for
details). Finally, we noticed that in a few cases the
extrapolation did not converge, probably due to an incompatibility
between the photospheric vector boundary condition and the
conditions at the outer boundaries of the computational volume. As
an example, the global extrapolation failed in case LL31; however,
when we limited the azimuthal integration region to $20^\circ <
\phi< 160^\circ$ the extrapolation was successful. The convergence
properties of our method certainly need further investigation.

The resulting solutions can be visually compared with the
corresponding benchmark solutions in Figure~1 where field lines
originating from the same photospheric positions are plotted. We
notice that the extrapolation produces field lines that are
visibly more stretched-out radially than the benchmark solution.
This is particularly obvious in the case of a current-free
displaced dipole and in the field lines that cross the outer
radial boundary at $r=2r_\odot$. This result is an artefact of our
method that qualitatively mimics the effect of the solar wind
beyond the outer radial distance.
%We remind the reader that the convergence
%parameters of the extrapolations to the benchmark solutions shown
%in Table~1 are calculated in a sub-region inside $r=1.5 r_\odot$.

\setcounter{table}{0}
\begin{table}
\begin{tabular}{llllllll}\hline
Case & $C_{\rm vec}$ & $C_{\rm CS}$ & $E'_n$ & $E'_m$ & CWsin &
$f$ & Comments \\ \hline DF & 0.989 & 0.986 & 0.845 & 0.821 &
\ldots & $10^{-9}$ &
Current-free\\
LL11 & 0.993 & 0.987 & 0.895 & 0.873 & 0.087 & $10^{-8}$ & as in
Tadesse
{\em et al.} (2009)\\
LL13 & 0.915 & 0.924 & 0.403 & 0.389 & 0.109 & $10^{-8}$ &
Different solution\\
LL31 & 0.981 & 0.978 & 0.772 & 0.715 & 0.110 & $10^{-7}$ &
Non-global in azimuth
\\
\hline \caption{Convergence parameters for various benchmark
extrapolations on a computational grid with a resolution in
($r,\theta,\phi$) of $10\times 34 \times 90$.}
\end{tabular}
\end{table}

\setcounter{figure}{0}
\begin{figure}[ht]
\begin{minipage}[b]{0.5\linewidth}
\centering
\includegraphics[trim=0cm 4cm 0cm 4cm,
clip=true, width=4.5cm, angle=270]{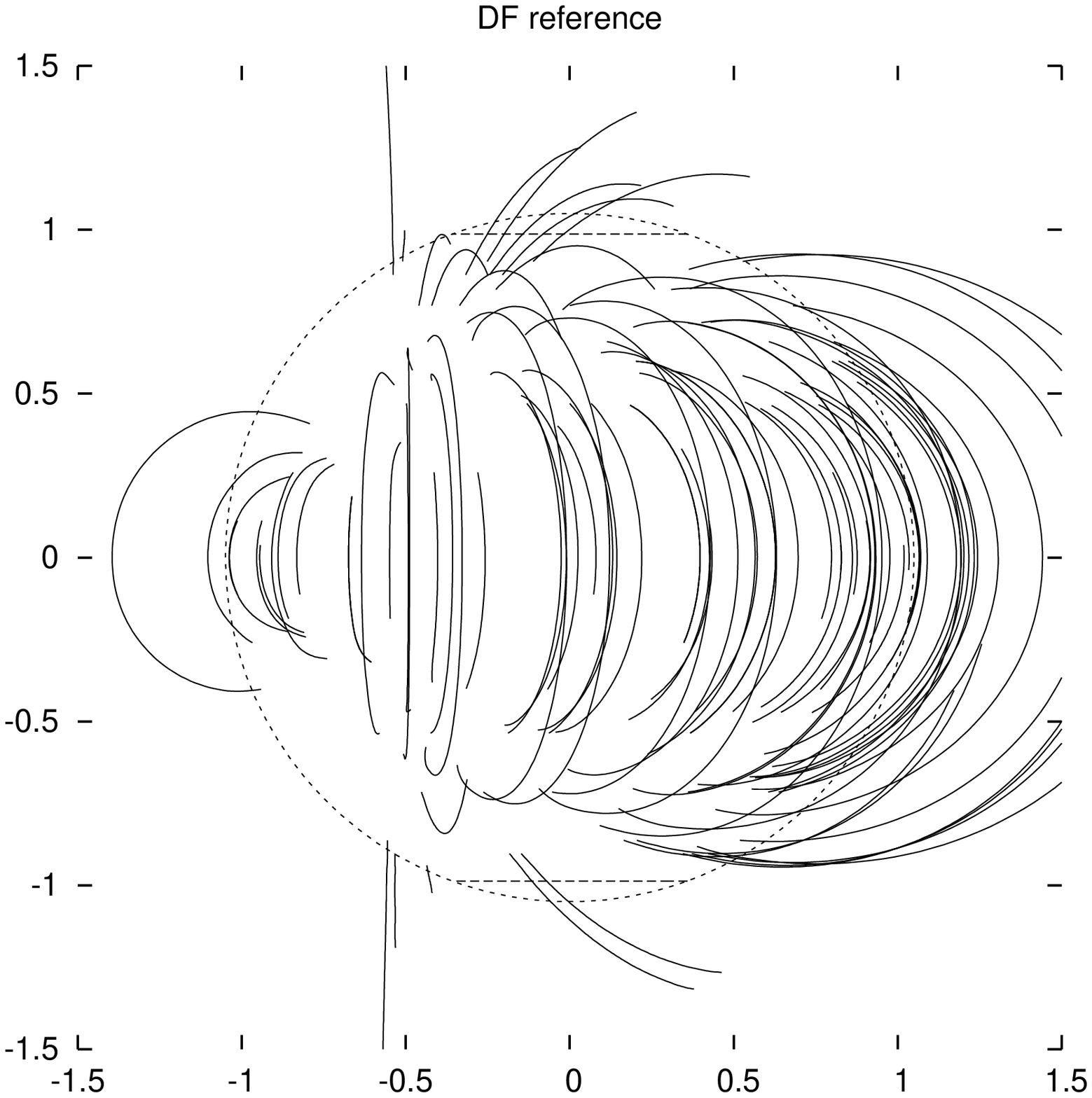}
%\caption{blah blah} \label{fig:blah1}
\end{minipage}
\hspace{1cm}
\begin{minipage}[b]{0.5\linewidth}
\centering
\includegraphics[trim=0cm 4cm 0cm 4cm,
clip=true, width=4.5cm, angle=270]{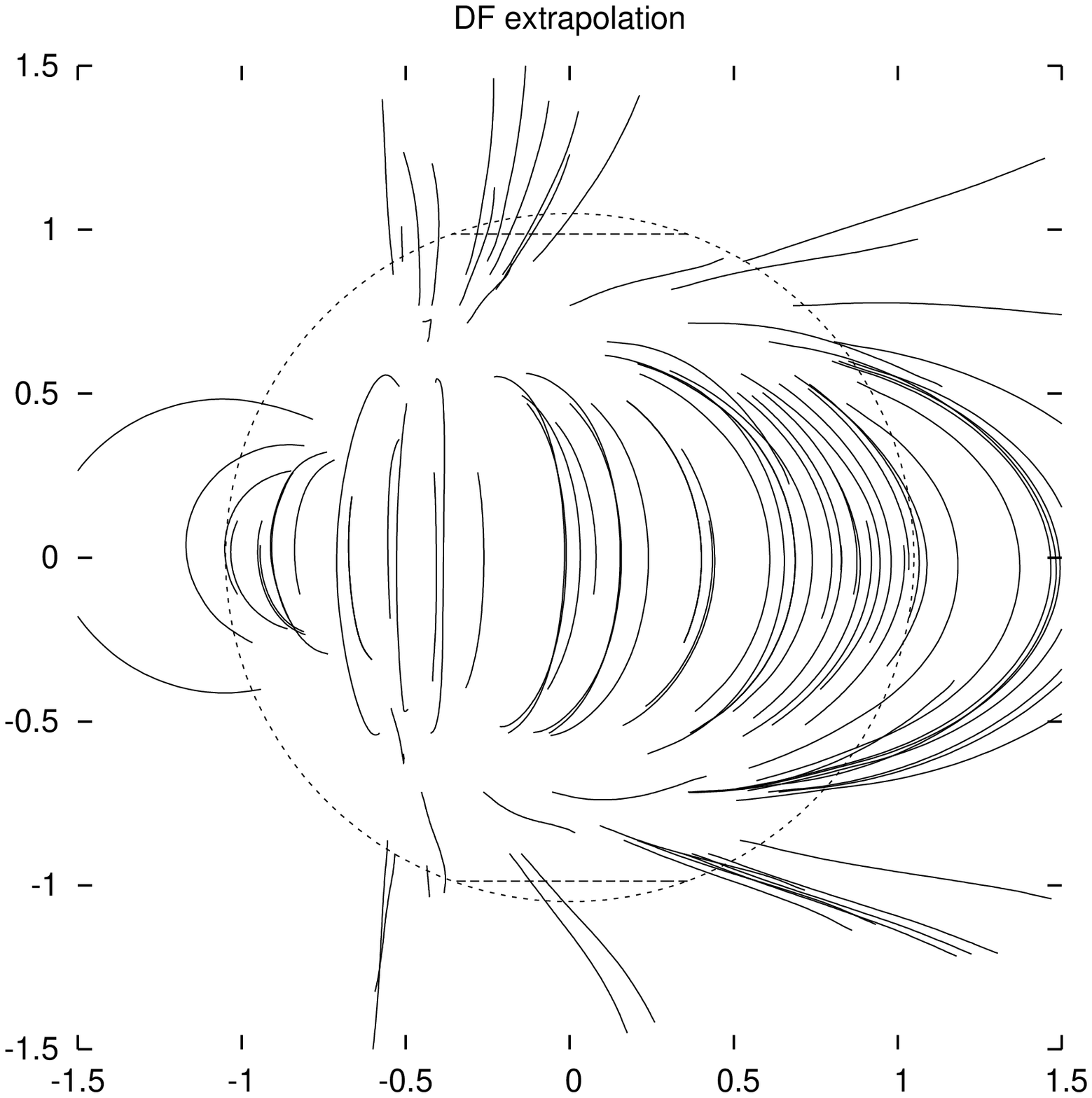}
%\caption{blah blah} \label{fig:blah2}
%  \includegraphics[width=4cm, angle=90]{L11.eps}
\end{minipage}
\begin{minipage}[b]{0.5\linewidth}
\centering
\includegraphics[trim=0cm 4cm 0cm 4cm,
clip=true, width=4.5cm, angle=270]{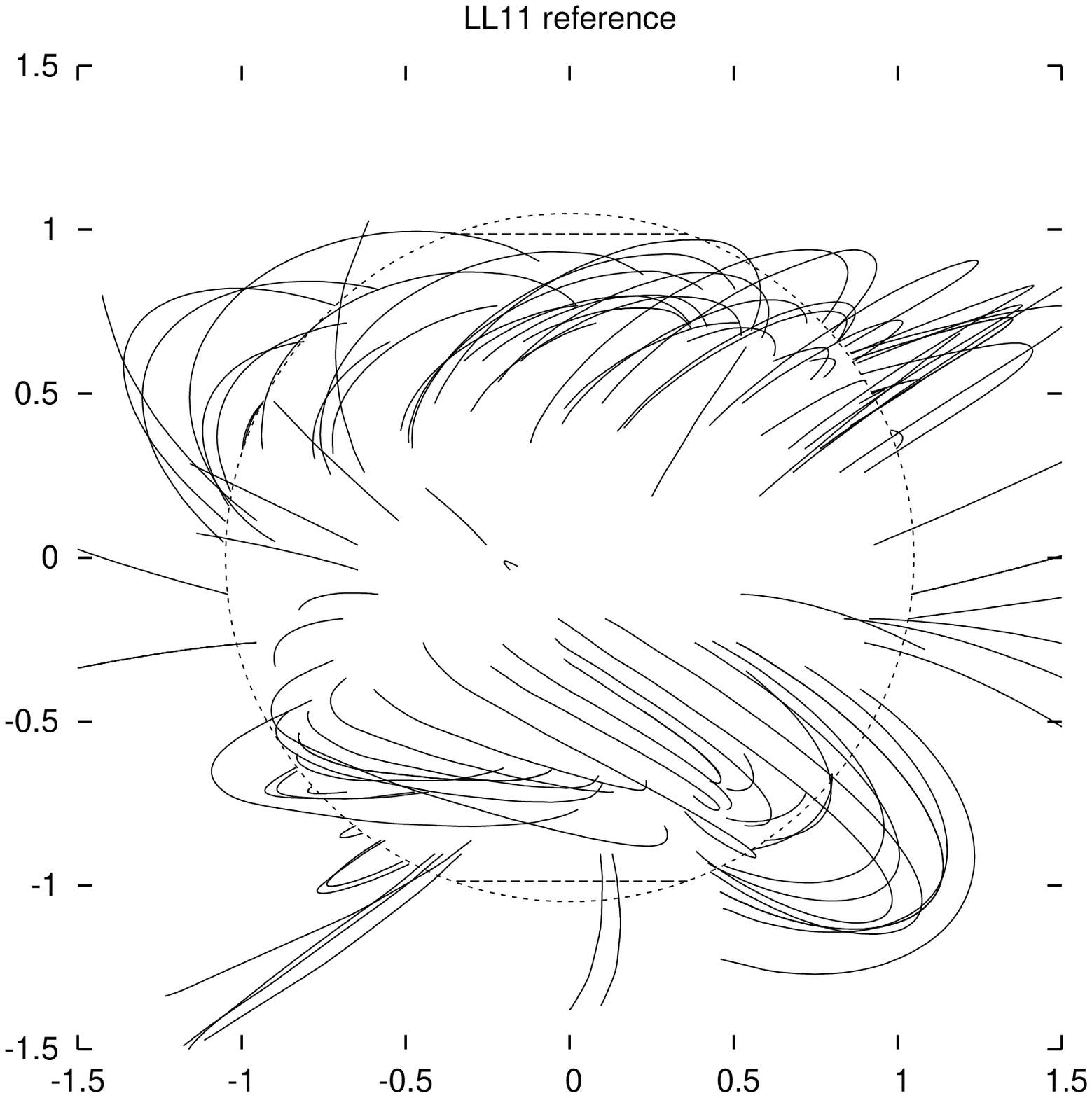}
%\caption{blah blah} \label{fig:blah1}
\end{minipage}
\hspace{1cm}
\begin{minipage}[b]{0.5\linewidth}
\centering
\includegraphics[trim=0cm 4cm 0cm 4cm,
clip=true, width=4.5cm, angle=270]{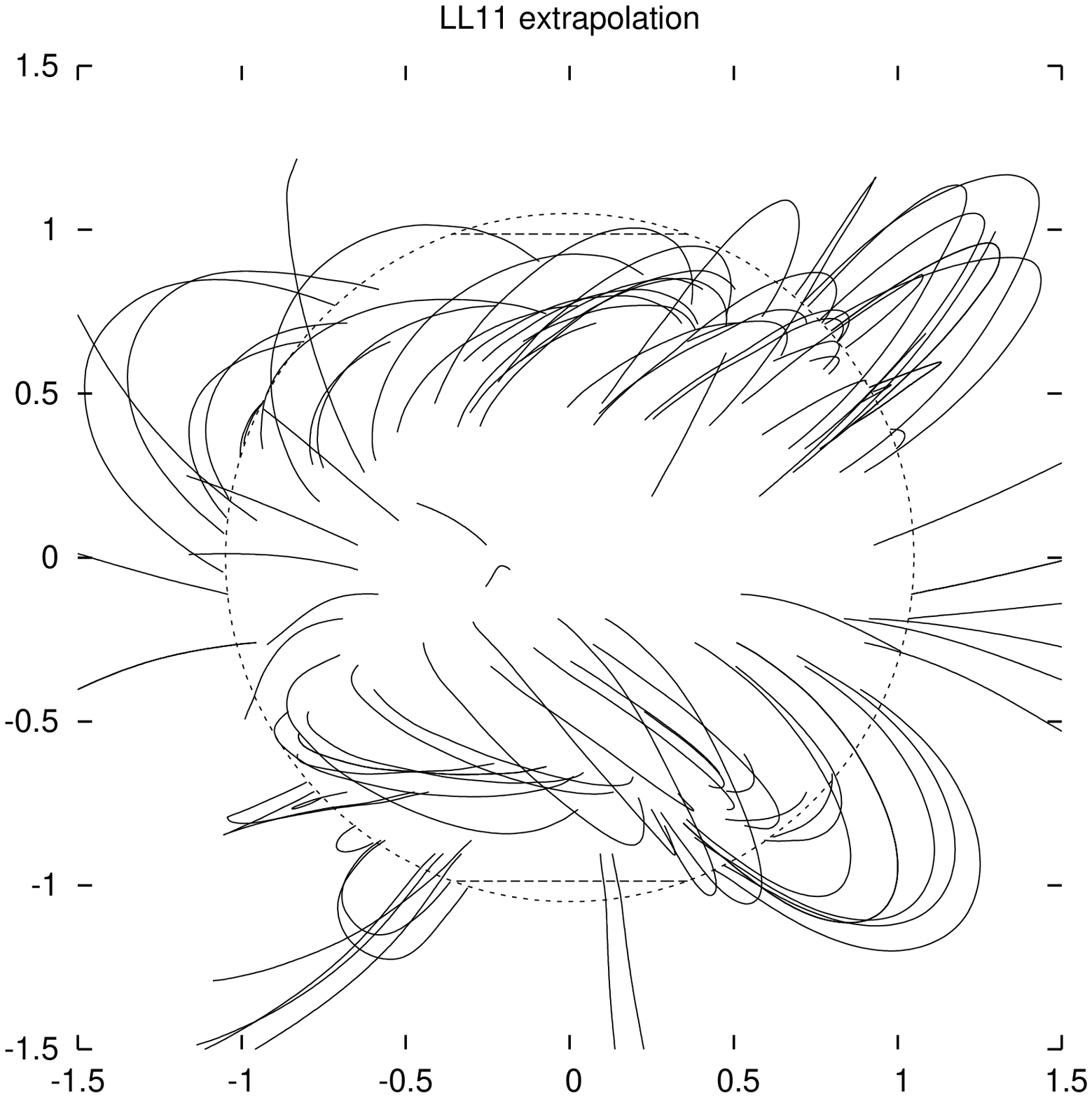}
%\caption{blah blah} \label{fig:blah2}
%  \includegraphics[width=4cm, angle=90]{L11.eps}
\end{minipage}
\begin{minipage}[b]{0.5\linewidth}
\centering
\includegraphics[trim=0cm 4cm 0cm 4cm,
clip=true, width=4.5cm, angle=270]{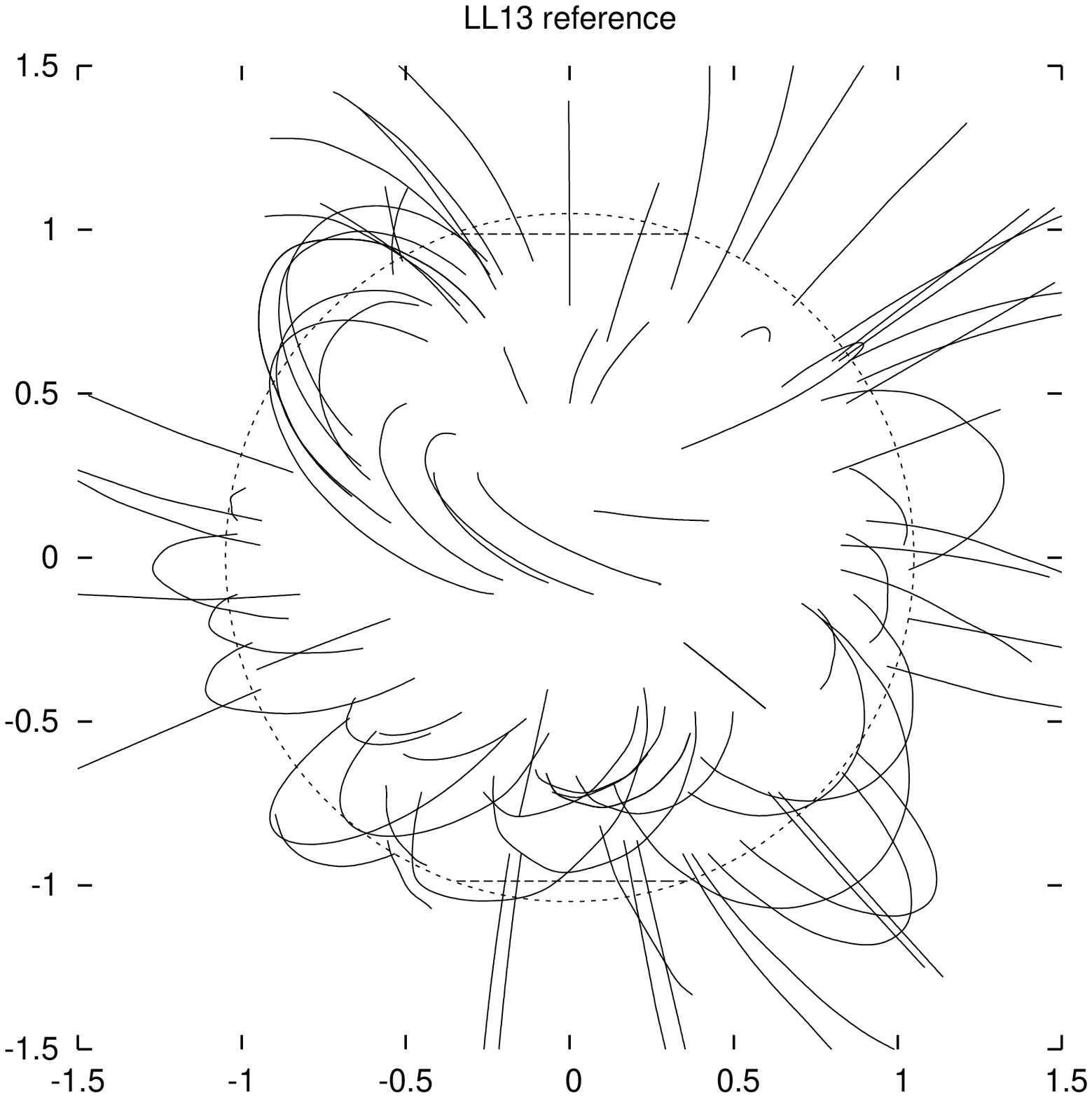}
%\caption{blah blah} \label{fig:blah1}
\end{minipage}
\hspace{1cm}
\begin{minipage}[b]{0.5\linewidth}
\centering
\includegraphics[trim=0cm 4cm 0cm 4cm,
clip=true, width=4.5cm, angle=270]{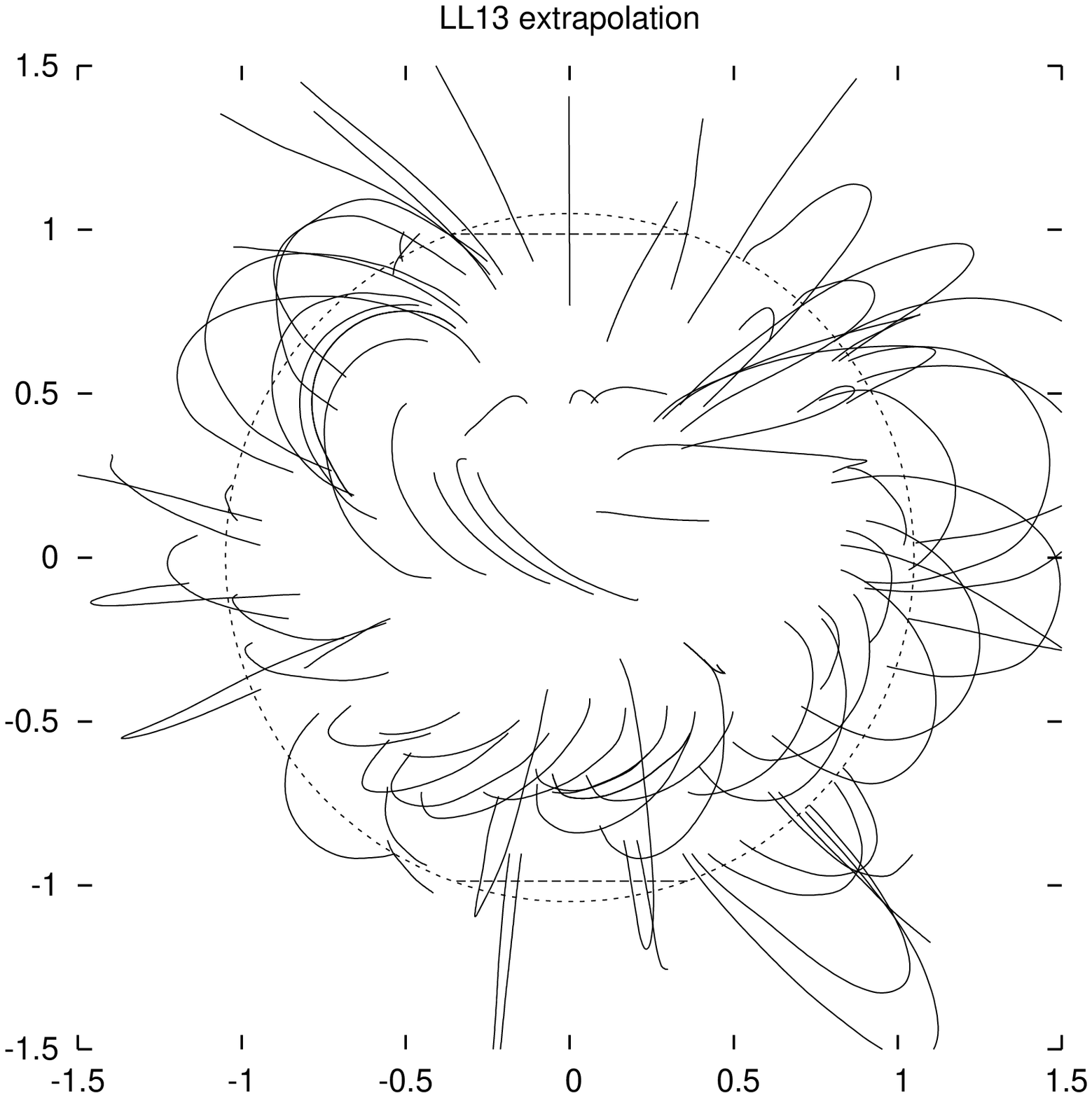}
%\caption{blah blah} \label{fig:blah2}
%  \includegraphics[width=4cm, angle=90]{L11.eps}
\end{minipage}
\begin{minipage}[b]{0.5\linewidth}
\centering
\includegraphics[trim=0cm 4cm 0cm 4cm,
clip=true, width=4.5cm, angle=270]{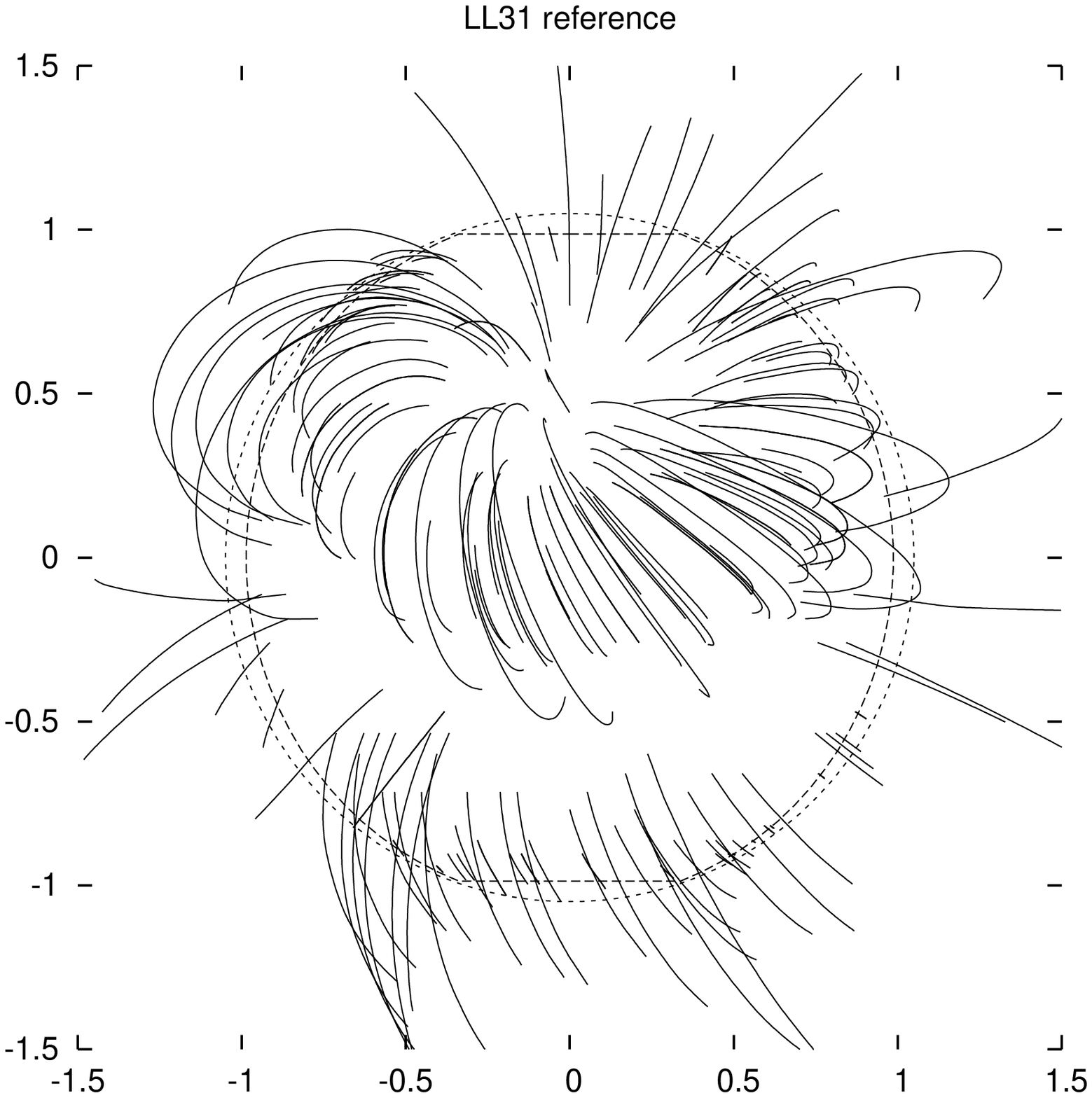}
%\caption{blah blah} \label{fig:blah1}
\end{minipage}
\hspace{1cm}
\begin{minipage}[b]{0.5\linewidth}
\centering
\includegraphics[trim=0cm 4cm 0cm 4cm,
clip=true, width=4.5cm, angle=270]{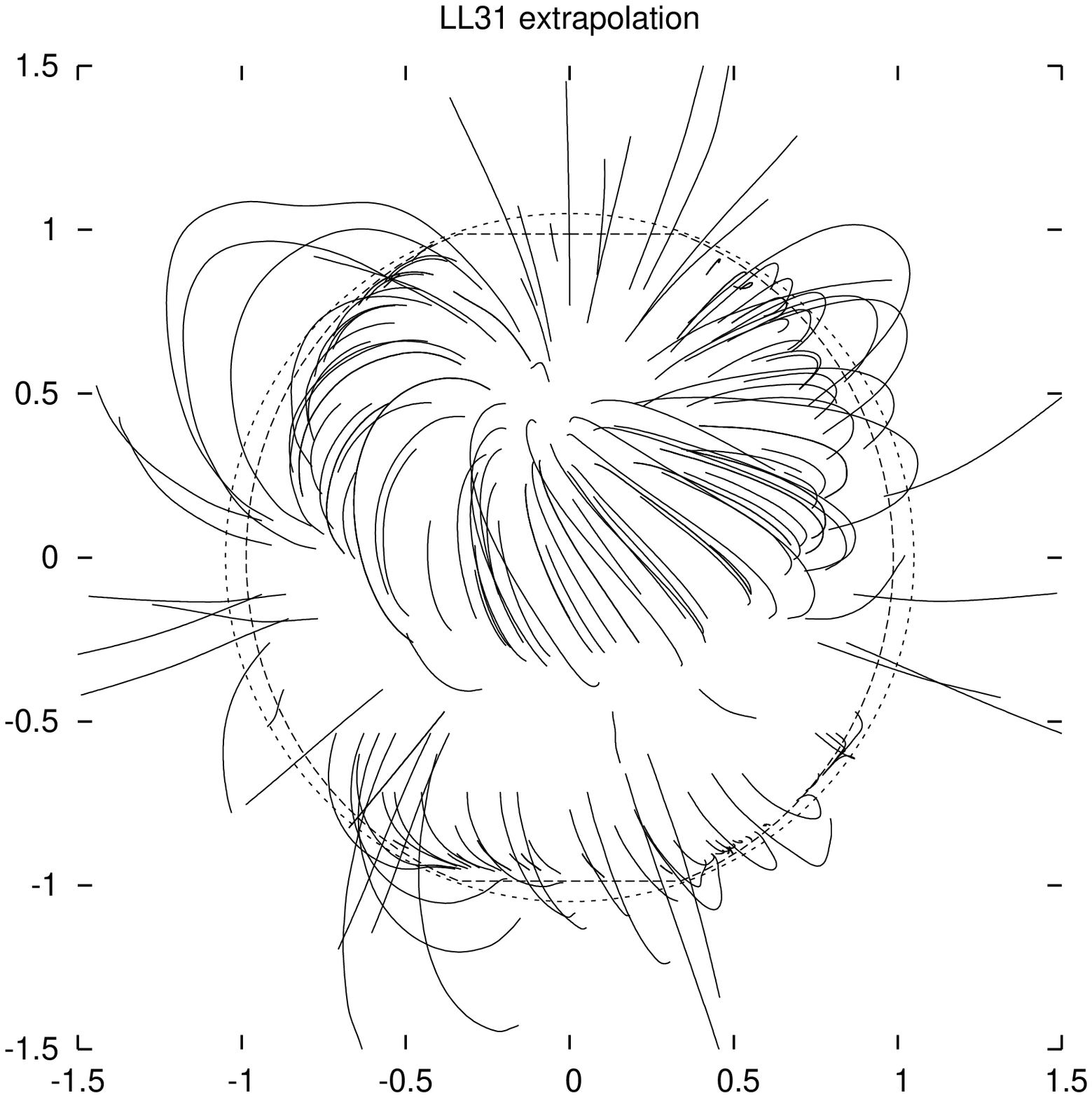}
%\caption{blah blah} \label{fig:blah2}
%  \includegraphics[width=4cm, angle=90]{L11.eps}
\end{minipage}
  \caption{Comparison between benchmark solutions (left)
  and their corresponding extrapolations (right) for the
  four cases of Table~1. In case LL11, the disk center corresponds to
  $180^\circ$ longitude. In the other three cases, it corresponds
  to $90^\circ$ longitude. Dashed lines mark the simulation
  photospheric boundaries.}
 \label{fig}
\end{figure}

\section{Prospects for the Future}

The implementation of the photospheric boundary condition from
full vector magnetograms in the new version of our code improves
the nonlinear force-free extrapolation of CKG in several respects.
Firstly, it generates a more physical solution (closer to
reality). Secondly, it resolves the non-uniqueness problem of the
original version where different initial conditions and different
photospheric evolutions yielded different magnetostatic
configurations. And thirdly, it does not allow the extrapolation
to `degenerate' to the potential configuration. Our test runs
suggest that the method is promising, as it outperforms existing
extrapolations in some benchmark cases. We therefore expect that
it has significant potential in the study of the lower solar
corona and of individual active regions.

We are currently running our code on a desktop PC. Our global
coronal magnetic field extrapolations were obtained with a
heliocentric angular resolution of $4^\circ\times 4^\circ$, on an
($r,\theta,\phi$) spherical computational grid of $10\times 34
\times 90$, but this can certainly be improved. Our goal is to
parallelize the code to run in a supercomputer. In a future
implementation we will also improve the performance of the PML
outer boundary by fine-tuning its parameters (our current
implementation allows for some minor amount of reflection of
transient electromagnetic-type waves).

An ambitious goal is to apply our method to investigate particular
active regions with minimum influence from the simulation outer
boundaries on the extrapolation. As we saw, the effect of the
outer boundaries differs from case to case, and we can only give
an empirical rule that the assumed volume must be about double the
size of the particular volume under investigation in all
directions (radial, polar, and azimuthal). Our goal for the future
is to solve for the magnetic field of a particular region as part
of ({\it i.e.} together with) a global coronal magnetic field
extrapolation. This can be achieved in practice by introducing
adaptive mesh refinement (AMR) around the region under
investigation and around neighboring regions that may interact
with that region. In practice, we will need about five adaption
levels in order to match the available heliocentric angular
resolution in particular active regions (about $0.05^\circ\times
0.05^\circ$) to the resolution that can be practically achieved on
a single PC in a global magnetic field extrapolation (about
$2^\circ\times 2^\circ$). Thus, the extrapolation for one
particular active region will take into account the presence of
neighboring active regions, as well as the global coronal magnetic
field structure.

One final goal is to introduce finite (non-infinite) conductivity
$\sigma$ in the code by modifying the FFE expression for the
electric current density as
\begin{equation}
{\bf J}=\rho_{\rm e}\frac{{\bf E}\times {\bf B}}{B^2}+\sigma {\bf
B}\ ,
\end{equation}
We plan to investigate various nonlinear prescriptions for
$\sigma$ ({\it e.g.} weighted by the value of $J$) as we are
currently doing in our study of the pulsar magnetosphere (see
\inlinecite{KKHC} for details). Finite conductivity will allow for
the presence of electric fields along the magnetic field, {\it
i.e.} it will introduce volume energy dissipation ${\bf J}\cdot
{\bf E}$. We expect that this approach will give us direct hints
about where and how magnetic field energy is released in the solar
corona.

%
%%%%%%%%%%%%%%%%%%%%%%%%%%%%%%%%%%%%%%%%%%%%%%%%%%%%%%%%%%%%%%%%%%%%%%%%%%%
\begin{acks}
This work was supported in part by the Research Commission of the
Academy of Athens.
\end{acks}

\appendix

\section*{Convergence Parameters}

We evaluate the following parameters introduced in
\inlinecite{Wetal}, \inlinecite{Setal}, and \inlinecite{Aetal}:
the vector correlation $C_{\rm vec}$
\begin{equation}
C_{\rm vec}\equiv \sum_i {\bf B}_i\cdot {\bf b}_i / \left( \sum_i
|{\bf B}_i|^2 \sum_i |{\bf b}_i|^2 \right)^{1/2}\ ,
\end{equation}
the parameter $C_{\rm CS}$
\begin{equation}
C_{\rm CS}\equiv \frac{1}{N}\sum_i \frac{{\bf B}_i\cdot {\bf
b}_i}{|{\bf B}_i||{\bf b}_i|}\ ,
\end{equation}
and the normalized and mean vector error $E'_n$ and $E'_m$
\begin{equation}
E'_n\equiv 1- \sum_i |{\bf B}_i - {\bf b}_i|/\sum_i |{\bf B}_i|\ ,
\end{equation}
\begin{equation}
E'_m\equiv 1- \frac{1}{N}\sum_i \frac{|{\bf B}_i-{\bf b}_i|}{|{\bf
B}_i|}\ ,
\end{equation}
%and the normalized total magnetic energy
%\begin{equation}
%\epsilon \equiv \frac{\Sum_i |{\bf b}_i|^2}{\Sum_i |{\bf B}_i|^2}\
%,
%\end{equation}
where ${\bf B}_i$ and ${\bf b}_i$ are the extrapolation and
reference fields, respectively, $i$ denotes grid points, and $N$
is the total number of grid points in the sub-region of our
simulation where we evaluate the convergence of the code. The
force-free and divergence-free conditions are estimated using the
following parameters:
\begin{equation}
\mbox{CWsin}\equiv \frac{\sum_i |{\bf J}_i\times {\bf B}_i|/|{\bf
B}_i|}{\sum_i|{\bf J}_i|}\ , \label{ff}
\end{equation}
\begin{equation}
f\equiv \frac{1}{N}\sum_i\frac{(\nabla\cdot {\bf B})_i}{6|{\bf
B}_i|/d}\ ,
\end{equation}
where $d$ is the minimum grid spacing. Those may be evaluated over
the total integration volume, or over a smaller internal volume.

%
%%% BIBLIOGRAPHY %%%%%%%%%%%%%%%%%%%%%%%%%%%%%%%%%%%%%%%%%%%%%%%%%%%%%%%%%%%
%
\bibliographystyle{spr-mp-sola}
\bibliography{SoPh_references}
\end{article}
\end{document}